# Pfcrmp May Play a Key Role in Chloroquine Antimalarial Action and Resistance Development

Gao-De Li*

School of Biological Sciences, University of Liverpool, Liverpool L69 7ZB, United Kingdom

**Abstract.** It was proposed earlier that Pfcrmp (*Plasmodium falciparum* chloroquine resistance marker protein) may be the chloroquine target protein in nucleus. In this communication, further evidence is presented to support the view that Pfcrmp may play a key role in chloroquine antimalarial actions as well as resistance development.

*Correspondence*: Dr. Gao-De Li, School of Biological Sciences, University of Liverpool, Liverpool L69 7ZB, United Kingdom. E-mail: gaode@liverpool.ac.uk

## 1. Introduction

The mechanism by which malaria parasites become chloroquine (CQ) resistant remains a mystery. Although Pfcrt K76T mutation has been widely applied as a marker in CQ resistance detection [1], the transporter's detailed role in CQ antimalarial action and resistance development is still not known. At present, many researches suggest that multiple genes may be involved in CQ resistance development [2-4].

It was hypothesized earlier that the nucleus may be the key site where CQ exerts its antimalarial actions, and later it was further proposed that Pfcrmp (*Plasmodium falciparum* chloroquine resistance protein) may be the target of CQ in the nucleus [5,6]. In this short communication, further evidence is presented to support the view that Pfcrmp may play a key role in CQ antimalarial action and resistance development.

## 2. Three genetic alterations in Pfcrmp gene are closely associated with CQR phenotypes

As described in a previous study, four deletion and two insertion mutations which are marked as D1, D2, D3, D4, In1, and In2, respectively, were found in CQ-resistant (CQR) *P. faciparum* K1 and Dd2 isolates, and none of them was seen in CQ-sensitive (CQS) *P. falciaprum* 3D7 and HB3 isolates [6]. Three genetic alterations of In1, D3, and D4 were selected as genetic markers for prediction of CQ resistance in *P. falciparum* because they were bigger in length than other genetic alterations and thus the corresponding PCR band-size differences between the wild-type and mutant type could be easily identified.

The 3 sets of PCR primer pairs which franked the three genetic alterations (listed in Table 1) were first used in checking the laboratory-maintained CQS HB3, CQR K1 and Dd2 isolates.

Any PCR band size that was different in length from that in control (3 PCR products obtained from HB3-isolate DNA were used as controls) was considered to be genetic alterations. The results showed that the 3 PCR band sizes were different in length among the 3 isolates (Fig. 1), which clearly indicated that the 3 genetic alterations can be examined directly by PCR.

Next the same PCR primers were used in detection of field isolates obtained from different regions including South East Asia, Africa and Papua New Guinea (PNG). The results showed that CQS isolates contained much less genetic alterations than CQR ones, and that the CQR isolates from different regions contained different numbers of genetic alterations (Table 2). Of the 29 CQR isolates from South East Asia, 22 isolates (75.9%) contained all 3 genetic alterations, 6 isolates (20.7%) contained 2 genetic alterations, and 1 isolate (3.4%) contained 1 genetic alteration; whereas Africa isolates showed a different pattern, i.e., of the 58 CQR isolates 2 isolates (3.4%) contained all 3 genetic alterations, 20 isolates (34.5%) contained 2 genetic alterations, and 35 isolates (60.3%) contained 1 genetic alteration. Furthermore, the PCR band sizes of the 3 genetic alterations in South East Asia isolates were either the same as those in Dd2 or K1 isolates, whereas a lot of the isolates from Africa exhibited new-type genetic alterations, their PCR band sizes were shorter or longer than those obtained from Dd2 and K1 isolates. Therefore, it is obvious that the genetic alteration distribution pattern in CQR isolates from South East Asia was different from that in CQR isolates from Africa, which is in agreement with the current opinion that CQ resistance in *P. falciparum* appears earlier and is more severe in South East Asia than in Africa.

**TABLE 1.** PCR primer pairs used in detection of 3 genetic alterations (In1, D3, and D4) in Pfcrmp gene.

| PCR primer pairs and cycling conditions | Genetic alteration examined by PCR | PCR band size in CQS HB3 | PCR band size indicating genetic alteration |
|---|---|---|---|
| 5'ATTATATATCAGATACATTATCGT3'<br>5'CTGCAGGAGTATAATTATTCAT3'<br>1st cycle: 94°C, 3 min; 55°C, 30 sec; 72°C, 40 sec.<br>38 cycles: 94°C, 30 sec; 53°C, 30 sec; 72°C, 40 sec | In1 | 300 bp | > 300 bp |
| 5'GAAAGATACAGCCAAGTATTATA3'<br>5'GTTTTCTTCATTACTTTGGTTTTTA3'<br>1st cycle: 94°C, 3 min; 57°C, 30 sec; 72°C, 1 min<br>38 cycles: 94°C, 30 sec; 55°C, 30 sec; 72°C, 1 min | D3 | 502 bp | <502 bp |
| 5'AGGATCAACAGGAAGATAAAG3'<br>5'CATGTACATGAAGTTGTTCGA3'<br>1st cycle: 94°C, 3 min; 55°C, 30 sec; 72°C, 1min<br>38 cycles: 94°C, 30 sec; 53°C, 30 sec; 72°C, 1min | D4 | 600 bp | < 600 bp |

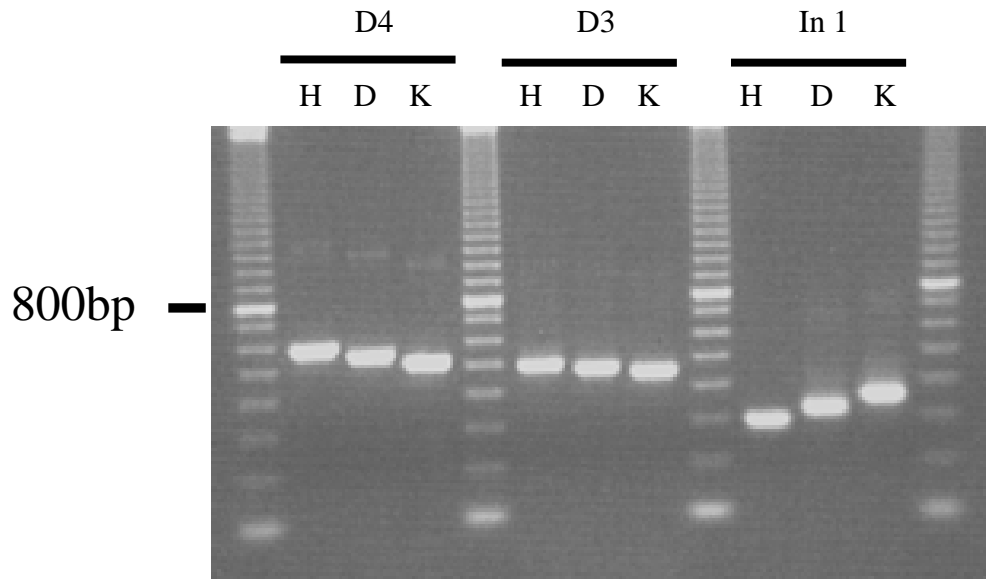

**Figure 1.** Three genetic alterations (In1, D3 and D4) of the Pfcrmp gene in CQS HB3 (H), CQR Dd2 (D), and K1 (K) isolates were examined by PCR. The PCR band sizes in D4: H = 600 bp, D = 564 bp, and K = 528 bp; in D3: H = 502 bp, D = 484 bp, and K = 466 bp; in In1: H = 300 bp, D = 330 bp, and K = 366 bp.

Based on the results that the 3 genetic alterations were closely associated with CQR phenotype, it is suggested that Pfcrmp may play a key role in CQ resistance development. The results also indicated that the 3 genetic alterations in Pfcrmp gene could be applied as markers in detection of CQ resistance in *P. falciparum*. The correct CQR-phenotype prediction rate could be 90.5% based on 1 genetic alteration, 97% based on 2 genetic alterations, and 100% based on 3 genetic alterations, respectively. Besides, this PCR-based method for detecting the 3 genetic alterations in Pfcrmp gene is much easier than that used in detection of point mutation (such as PfCRT K76T mutation) in which DNA sequencing or restriction enzyme digestion is needed.

## 3. Pfcrmp's Function and its Linkage to CQ Antimalarial Actions

To date, Pfcrmp's function remains unknown because its protein sequence does not appear to share good homology with many other known protein sequences. Nevertheless, after protein sequence analyses using servers of PredictProtein (www.predictprotein.org) and ExPASy (http://expasy.org/tools/), some clues indicating Pfcrmp's functions were obtained.

**TABLE 2. CQR phenotype and the 3 genetic alterations (In1, D3 and D4) of Pfcrmp gene in isolates from different regions.**

|  |  |  | Number of the 3 genetic alterations in each isolate | | | |
|---|---|---|---|---|---|---|
|  | Regions | Total No. | 0 | 1 | 2 | 3 |
| CQS isolates ($IC_{50}$ ≤21 nM) | All regions | 18 | 13 | 4 | 1 | 0 |
| CQR isolates ($IC_{50}$ ≥22 nM) | S.E. Asia | 29 | 0 | 1 | 6 | 22 |
|  | Africa | 58 | 1 | 35 | 20 | 2 |
|  | PNG | 8 | 0 | 2 | 6 | 0 |
| % of CQR phenotype | – | – | 7.1 (1/14) | 90.5 (38/42) | 97.0 (32/33) | 100.0 (24/24) |

First of all, it is confirmed by the analyses that Pfcrmp is a DNA-binding nuclear protein for it contains DNA binding signals and several nuclear localization signals. Secondly, Pfcrmp might be involved in transcriptional regulation for it contains many phosphorylation sites (mainly PKC), and 5 nine amino acid transactivation domains [7]. Thirdly, PSI-BLAST multiple sequence alignment showed that 24% of pairwise sequence identity and 43% of similarity were obtained after alignment of 211 amino acid residues between Pfcrmp and *P. falciparum* flap endonuclease 1 (AF093702), suggesting that Pfcrmp may have nuclease activity, and probably is involved in DNA repair. Fourthly, InterProScan analysis showed that in the C-terminal amino acids 3141-3210 of Pfcrmp there is a domain that contains 5 cysteine residues and is similar to insect cysteine-rich antifreeze protein [8]. Since the antifreeze-like domain has been suggested to be important for substrate binding in human sialic acid synthase [9], and that CQ can bind to certain sulfhydryl groups of human serum protein [10], it is possible that CQ can bind to the antifreeze-like domain region of Pfcrmp. Finally, myosin II heavy chain has been used as template PDB to build Pfcrmp's model (ModBase Q968T7), suggesting that Pfcrmp is structurally similar to myosin II heavy chain. Since myosin molecule can walk along an actin filament, perhaps, Pfcrmp could walk along DNA strand.

Taken together, it is reasonable to speculate that Pfcrmp may function as a DNA-binding nuclear protein that could walk along the DNA strand for repairing errors and may also be involved in other important functions such as transcriptional regulation. The CQ antimalarial action may be linked to Pfcrmp's function. It was hypothesized that in CQS isolates CQ may bind to wild-type Pfcrmp or intercalate into the DNA region to which the wild-type Pfcrmp happens to bind, and then interact with each other, as a result the wild-type Pfcrmp's nuclease activity is abnormally activated, causing DNA damage which could lead to parasite death through p53-dependent apoptosis pathway [11].

In CQR isolates, probably due to conformational alteration caused by insertions and deletions, the mutant Pfcrmp could not interact with CQ or interact with CQ in a way different from its wild-type counterpart's, and thus prevent CQ from damaging DNA. A recently published paper suggests that malaria parasite employs gene duplications and deletions as general strategies to enhance its survival [12].

As mentioned above, Pfcmp may play a very important role in malaria parasite survival and spread. Mutant Pfcrmp may help CQR parasite to survive under CQ pressure, but it may weaken CQR parasite's ability to compete with CQS parasite for survival after CQ withdrawal. This could be used in explanation of recovery of CQ sensitivity after cessation of CQ use in some regions [13-15]. Besides, impaired functions of Pfcrmp may impair genome stability and increase mutation rate, which can be used to explain the fact that once malaria parasite becomes CQR its resistance to other antimalarial drugs including structurally similar and non-similar ones will be easily developed.

## 4. Conclusion

Based on the preliminary experimental data that the 3 genetic alterations in Pfcrmp gene are closely linked to CQR phenotype, it is suggested that Pfcrmp may be a key player in CQ resistance development. Protein sequence analyses suggest that Pfcrmp is a DNA-binding nuclear protein that may be involved in DNA repair and other important functions. CQ antimalarial action may be linked to Pfcrmp's function, and further investigation into detailed CQ-Pfcrmp interaction is needed.